\documentclass[a4paper,11pt]{article}

\usepackage[top=1.0in,right=1.0in,left=1.0in,bottom=1.75in]{geometry}
\usepackage{latexsym}
\usepackage{amssymb}
\usepackage{amsmath,amsfonts,amssymb,amsthm}
\usepackage{float}
\usepackage{graphics,graphicx}
\usepackage{tabularx}
\usepackage{amsfonts}
\usepackage{amsmath}
\usepackage{enumerate}
\usepackage{booktabs}
\usepackage{multirow}
\usepackage{sectsty}
\usepackage[affil-it,auth-sc]{authblk}
\usepackage{subcaption}
\usepackage[usenames,dvipsnames]{xcolor}

\usepackage{tabu}

\usepackage{fancyhdr}            % NB: E=pagina pari, O=pagina dispari, H=intestazione, F=piede pagina, L,R,C=sinistra,destra,centro ; \leftmark = capitolo ; \rightmark = paragrafo
\fancypagestyle{plain}{%
  \fancyhead{}                                     % Pulisce il campo della testata
  \fancyfoot[CE,CO]{\thepage}       % Numero pagina in basso ed esterno in grassetto
  \fancyhead[CE,CO]{\textcolor[rgb]{0.64,0.15,0.15}{Published in \emph{International Journal of Solids and Structures} (2015) \textbf{69-70} 491-497
\\
doi: http://dx.doi.org/10.1016/j.ijsolstr.2015.05.004}}
\setlength{\headheight}{14.5pt}}
\sectionfont{\normalsize}
\subsectionfont{\normalsize}
\begin{document}

%definitions: general
\newcommand{\singlespace}{\baselineskip=12pt\lineskiplimit=0pt\lineskip=0pt}
\def\ds{\displaystyle}

%definitions: equations
\newcommand{\beq}{\begin{equation}}
\newcommand{\eeq}{\end{equation}}
\newcommand{\lb}{\label}
\newcommand{\ph}{\phantom}
\newcommand{\beqar}{\begin{eqnarray}}
\newcommand{\eeqar}{\end{eqnarray}}
\newcommand{\barr}{\begin{array}}
\newcommand{\earr}{\end{array}}
\newcommand{\jump}{\parallel}
\newcommand{\Ehat}{\hat{E}}
\newcommand{\That}{\hat{\bf T}}
\newcommand{\Ahat}{\hat{A}}
\newcommand{\chat}{\hat{c}}
\newcommand{\shat}{\hat{s}}
\newcommand{\khat}{\hat{k}}
\newcommand{\muhat}{\hat{\mu}}
\newcommand{\mc}{M^{\scriptscriptstyle C}}
\newcommand{\mei}{M^{\scriptscriptstyle M,EI}}
\newcommand{\mec}{M^{\scriptscriptstyle M,EC}}
\newcommand{\hbeta}{{\hat{\beta}}}
\newcommand{\rec}[2]{\left( #1 #2 \ds{\frac{1}{#1}}\right)}
\newcommand{\rep}[2]{\left( {#1}^2 #2 \ds{\frac{1}{{#1}^2}}\right)}
\newcommand{\derp}[2]{\ds{\frac {\partial #1}{\partial #2}}}
\newcommand{\derpn}[3]{\ds{\frac {\partial^{#3}#1}{\partial #2^{#3}}}}
\newcommand{\dert}[2]{\ds{\frac {d #1}{d #2}}}
\newcommand{\dertn}[3]{\ds{\frac {d^{#3} #1}{d #2^{#3}}}}
\newcommand{\ct}{\captionof{table}}
\newcommand{\cf}{\captionof{figure}}
\newcommand{\dd}{\diff}
\newcommand{\rr}{\textcolor{red}}
\def\c{{\circ}}
\def\bob{{\, \underline{\overline{\otimes}} \,}}
\def\ob{{\, \underline{\otimes} \,}}
\def\scalp{\mbox{\boldmath$\, \cdot \, $}}
\def\gdp{\makebox{\raisebox{-.215ex}{$\Box$}\hspace{-.778em}$\times$}}
\def\daa{\makebox{\raisebox{-.050ex}{$-$}\hspace{-.550em}$: ~$}}
\def\mK{\mbox{${\mathcal{K}}$}}
\def\cK{\mbox{${\mathbb {K}}$}}

%definitions: integrals
\def\Xint#1{\mathchoice
   {\XXint\displaystyle\textstyle{#1}}%
   {\XXint\textstyle\scriptstyle{#1}}%
   {\XXint\scriptstyle\scriptscriptstyle{#1}}%
   {\XXint\scriptscriptstyle\scriptscriptstyle{#1}}%
   \!\int}
\def\XXint#1#2#3{{\setbox0=\hbox{$#1{#2#3}{\int}$}
     \vcenter{\hbox{$#2#3$}}\kern-.5\wd0}}
\def\ddashint{\Xint=}
\def\fpint{\Xint=}
\def\dashint{\Xint-}
\def\cpvint{\Xint-}
\def\intl{\int\limits}
\def\cpvintl{\cpvint\limits}
\def\fpintl{\fpint\limits}
\def\ointl{\oint\limits}
\def\bA{{\bf A}}
\def\ba{{\bf a}}
\def\bB{{\bf B}}
\def\bb{{\bf b}}
\def\bc{{\bf c}}
\def\bC{{\bf C}}
\def\bD{{\bf D}}
\def\bE{{\bf E}}
\def\be{{\bf e}}
\def\bbf{{\bf f}}
\def\bF{{\bf F}}
\def\bG{{\bf G}}
\def\bg{{\bf g}}
\def\bi{{\bf i}}
\def\bH{{\bf H}}
\def\bK{{\bf K}}
\def\bL{{\bf L}}
\def\bM{{\bf M}}
\def\bN{{\bf N}}
\def\bn{{\bf n}}
\def\bm{{\bf m}}
\def\b0{{\bf 0}}
\def\bo{{\bf o}}
\def\bX{{\bf X}}
\def\bx{{\bf x}}
\def\bP{{\bf P}}
\def\bp{{\bf p}}
\def\bQ{{\bf Q}}
\def\bq{{\bf q}}
\def\bR{{\bf R}}
\def\bS{{\bf S}}
\def\bs{{\bf s}}
\def\bT{{\bf T}}
\def\bt{{\bf t}}
\def\bU{{\bf U}}
\def\bu{{\bf u}}
\def\bv{{\bf v}}
\def\bw{{\bf w}}
\def\bW{{\bf W}}
\def\by{{\bf y}}
\def\bz{{\bf z}}
\def\T{{\bf T}}
\def\Te{\textrm{T}}
\def\Id{{\bf I}}
\def\bxi{\mbox{\boldmath${\xi}$}}
\def\balpha{\mbox{\boldmath${\alpha}$}}
\def\bbeta{\mbox{\boldmath${\beta}$}}
\def\bepsilon{\mbox{\boldmath${\epsilon}$}}
\def\bvarepsilon{\mbox{\boldmath${\varepsilon}$}}
\def\bomega{\mbox{\boldmath${\omega}$}}
\def\bphi{\mbox{\boldmath${\phi}$}}
\def\bsigma{\mbox{\boldmath${\sigma}$}}
\def\bfeta{\mbox{\boldmath${\eta}$}}
\def\bDelta{\mbox{\boldmath${\Delta}$}}
\def\btau{\mbox{\boldmath $\tau$}}
\def\tr{{\rm tr}}
\def\dev{{\rm dev}}
\def\div{{\rm div}}
\def\Div{{\rm Div}}
\def\Grad{{\rm Grad}}
\def\grad{{\rm grad}}
\def\Lin{{\rm Lin}}
\def\Sym{{\rm Sym}}
\def\Skw{{\rm Skew}}
\def\abs{{\rm abs}}
\def\Re{{\rm Re}}
\def\Im{{\rm Im}}
\def\capB{\mbox{\boldmath${\mathsf B}$}}
\def\capC{\mbox{\boldmath${\mathsf C}$}}
\def\capD{\mbox{\boldmath${\mathsf D}$}}
\def\capE{\mbox{\boldmath${\mathsf E}$}}
\def\capG{\mbox{\boldmath${\mathsf G}$}}
\def\tcapG{\tilde{\capG}}
\def\capH{\mbox{\boldmath${\mathsf H}$}}
\def\capK{\mbox{\boldmath${\mathsf K}$}}
\def\capL{\mbox{\boldmath${\mathsf L}$}}
\def\capM{\mbox{\boldmath${\mathsf M}$}}
\def\capR{\mbox{\boldmath${\mathsf R}$}}
\def\capW{\mbox{\boldmath${\mathsf W}$}}

%imaginary unit
\def\i{\mbox{${\mathrm i}$}}
\def\mC{\mbox{\boldmath${\mathcal C}$}}
\def\mB{\mbox{${\mathcal B}$}}
\def\mE{\mbox{${\mathcal{E}}$}}
\def\mL{\mbox{${\mathcal{L}}$}}
\def\mK{\mbox{${\mathcal{K}}$}}
\def\mV{\mbox{${\mathcal{V}}$}}
\def\C{\mbox{\boldmath${\mathcal C}$}}
\def\E{\mbox{\boldmath${\mathcal E}$}}

%definitions: journals
\def\AAM{{\it Advances in Applied Mechanics }}
\def\ACME{{\it Arch. Comput. Meth. Engng.}}
\def\ARMA{{\it Arch. Rat. Mech. Analysis}}
\def\AMR{{\it Appl. Mech. Rev.}}
\def\ASCEEM{{\it ASCE J. Eng. Mech.}}
\def\ACTA{{\it Acta Mater.}}
\def\CMAME {{\it Comput. Meth. Appl. Mech. Engrg.}}
\def\CRAS{{\it C. R. Acad. Sci. Paris}}
\def\CRM{{\it Comptes Rendus M\'ecanique}}
\def\EFM{{\it Eng. Fracture Mechanics}}
\def\EJMA{{\it Eur.~J.~Mechanics-A/Solids}}
\def\IJES{{\it Int. J. Eng. Sci.}}
\def\IJF{{\it Int. J. Fracture}}
\def\IJMS{{\it Int. J. Mech. Sci.}}
\def\IJNAMG{{\it Int. J. Numer. Anal. Meth. Geomech.}}
\def\IJP{{\it Int. J. Plasticity}}
\def\IJSS{{\it Int. J. Solids Structures}}
\def\IngA{{\it Ing. Archiv}}
\def\JAM{{\it J. Appl. Mech.}}
\def\JAP{{\it J. Appl. Phys.}}
\def\JAE{{\it J. Aerospace Eng.}}
\def\JE{{\it J. Elasticity}}
\def\JM{{\it J. de M\'ecanique}}
\def\JMPS{{\it J. Mech. Phys. Solids}}
\def\JSV{{\it J. Sound and Vibration}}
\def\MACRO{{\it Macromolecules}}
\def\MMT{{\it Mech. Mach. Th.}}
\def\MOM{{\it Mech. Materials}}
\def\MMS{{\it Math. Mech. Solids}}
\def\MMT{{\it Metall. Mater. Trans. A}}
\def\MPCPS{{\it Math. Proc. Camb. Phil. Soc.}}
\def\MSE{{\it Mater. Sci. Eng.}}
\def\NATURE{{\it Nature}}
\def\NATUREM{{\it Nature Mater.}}
\def\PHIL{{\it Phil. Trans. R. Soc.}}
\def\PMPS{{\it Proc. Math. Phys. Soc.}}
\def\PNAS{{\it Proc. Nat. Acad. Sci.}}
\def\PRE{{\it Phys. Rev. E}}
\def\PRL{{\it Phys. Rev. Letters}}
\def\PRSL{{\it Proc. R. Soc.}}
\def\RIIT{{\it Rozprawy Inzynierskie - Engineering Transactions}}
\def\ROCK{{\it Rock Mech. and Rock Eng.}}
\def\QAM{{\it Quart. Appl. Math.}}
\def\QJMAM{{\it Quart. J. Mech. Appl. Math.}}
\def\SCIENCE{{\it Science}}
\def\SCRMAT{{\it Scripta Mater.}}
\def\SM{{\it Scripta Metall.}}
\def\ZAMM{{\it Z. Angew. Math. Mech.}}
\def\ZAMP{{\it Z. Angew. Math. Phys.}}
\def\ZVDI{{\it Z. Verein. Deut. Ing.}}
\renewcommand\Affilfont{\itshape\small}
\setlength{\affilsep}{1em}
\renewcommand\Authsep{, }
\renewcommand\Authand{ and }
\renewcommand\Authands{ and }
\setcounter{Maxaffil}{3}
\definecolor{traz}{RGB}{190,30,0}
\definecolor{comp}{RGB}{0,50,190}

\title{
The deformation of an elastic rod with a clamp \\ sliding along a smooth and curved profile
}
\author[1]{D. Misseroni}
\author[2]{G. Noselli}
\author[3]{D. Zaccaria}
\author[1]{D. Bigoni\footnote{Corresponding author. Phone:\,+39\,0461\,282507; E-mail:\,bigoni@ing.unitn.it; Fax:\,+39\,0461\,282599.}}
\affil[1]{DICAM, University of Trento, via~Mesiano~77, I-38123 Trento, Italy.}
\affil[2]{SISSA--International School for Advanced Studies, via~Bonomea~265, I-34136 Trieste, Italy.}
\affil[3]{DICAr, University of Trieste, piazzale~Europa~1, I-34127 Trieste, Italy.}

\date{}
\maketitle

\begin{abstract}

\noindent
The design of compliant mechanisms is crucial in several technologies and relies on the availability of solutions for nonlinear structural problems. One of these solutions is given and experimentally validated 
in the present article for a compliant mechanism moving along a smooth curved profile. In particular, a deformable elastic rod is held by two clamps, one at each end. The first clamp is constrained to slide without friction along a curved profile, while the second clamp moves in a straight line transmitting its motion through the elastic rod to the first clamp. 
For this system it is shown that the clamp
sliding on the profile imposes nontrivial boundary conditions (derived via a variational and an
asymptotic approach), which strongly influence buckling and nonlinear structural behaviour.
Investigation of this behaviour shows that a compliant mechanism can be designed, which gives an almost neutral response in compression. This behavior could easily be exploited to make a force limiting device. Finally a proof-of-concept device was constructed and tested showing that the analyzed mechanical system can be realized in practice and it behaves tightly to the model, so that it can now be used in the design of machines that use compliant mechanisms. 

\end{abstract}


\section{Introduction}\lb{S_INTRO}

Compliant mechanisms are going through a paradigm change. Where once they were part of the fixtures and fittings of mechanisms they are becoming the mechanisms themselves. This is especially true in nano- and 
micro- mechanics where joints, linkages and their associated bearings are difficult to make and assemble. They are also important in bioinspired systems, as biological systems are often composed of soft elements working over a large range of displacements. Advances in this field are heavily reliant on the available solutions for the non linear behaviour of structural elements as well as physical proof that the theoretical models can be realized in practice. 
The purpose of the present article is to investigate, both theoretically and experimentally, the planar elastica with a \lq non-standard constraint' applied at one of its ends, namely, a clamp that is constrained to follow a curved and frictionless profile. The influence of  constraints of this type on elastic rods has been recently highlighted by Zaccaria {\it et al.} (2011), showing that a slider can introduce tensile buckling in an elastic system\footnote{
Bifurcation for tensile loads was also addressed by Ziegler (1977), but a compressed element responsible of buckling is present in his example, and by Gajewski and Palej (1974), as a result of a live 
load. Zyczkowski (1991) concludes that under dead loading buckling is impossible when all structural elements are subject to tension. Biezeno and Grammel (1955) fail to notice that one structure 
considered by them as an example of multiple loadings displays tensile buckling, when subjected to a certain load. The example reported by Zaccaria {\it et al.} (2011) shows that tensile buckling of a 
system in which all elements are subject to tension is possible under a dead load. To the best of the authors' knowledge, and excluding a situation analyzed by Timoshenko and Gere (1936), the effect of constraint curvature 
on buckling load was only considered by Bigoni {\it et al.} (2012, 2013).}, and by Bigoni {\it et al.} (2012, 2013), demonstrating the strong influence of constraint curvature on buckling (which, by \lq playing' with the sign of the curvature, may be turned from compressive to tensile) and postcritical behaviour.

\begin{figure}[!htcb]
\renewcommand{\figurename}{\footnotesize{Fig.}}
\begin{center}
\includegraphics[width=12.5 cm]{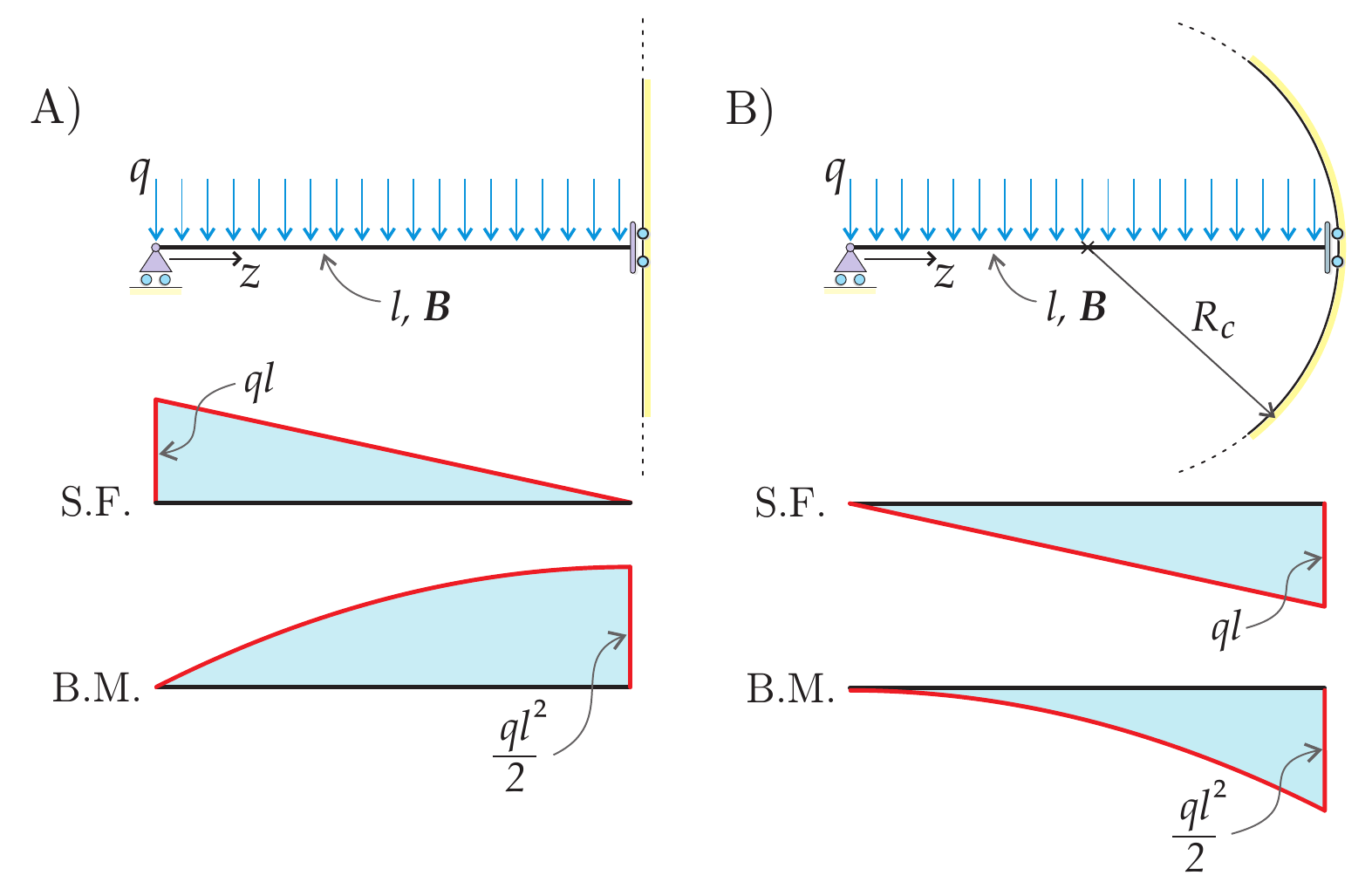}
\caption{\footnotesize
Two elastic structures (the rods have a length $l$ and are subject to a vertical, distributed load $q$) differing only in the curvature of the profile along which the clamp on the right end can slide (a vertical line on the left, a circle of radius $R_c=l/2$ on the right) exhibit a completely different mechanical behaviour, so that, while the structure on the left is equivalent to one half of a simply-supported beam,
that on the right is equivalent to a beam simply clamped on the right and free at the other end. 
\lq S.F.' and \lq B.M.' stand for \lq shear force' and \lq bending moment', respectively.
}
\label{fig:intro}
\end{center}
\end{figure}

To illustrate the effect of a \lq non-standard constraint' on structural systems, let us consider the two simple structures of length $l$ shown in Fig.\,\ref{fig:intro}, both subjected to a distributed transverse  load of magnitude $q$ and simply supported at the left end (the rods are assumed inextensible and solved in the small deflection approximation). The difference between the two structures lies in the constraint applied on their right end: 
figure A shows a clamp that is free to slide along a vertical line while figure B shows a clamp that is constrained to follow a curve, 
which for simplicity is circular with $R_c = l/2$.
%%%%%%%%%% FIG %%%%%%%%%%
%%%%%%%%%% END FIG %%%%%%%%%%
It can be appreciated from the diagrams of the bending moment (B.M.) and of the shear force (S.F.) shown in the figure that the two elastic solutions are totally different, a fact that demonstrates the strong influence of the constraint's curvature.

In order to derive the natural boundary condition that emerges from the presence of the curved constraint, it suffices to write the total potential energy of the elastic system shown in Fig.\,\ref{fig:intro}, expressed as a function of the transverse displacement $v(z)$,
\begin{equation}
\label{eq:EPT}
\mathcal{V}(v) = \frac{1}{2} B \intop_{0}^{l} \left(\frac{d^2 v(z)}{dz^2}\right)^2 \, dz - q  \intop_{0}^{l} v(z) \, dz ,
\end{equation}
and to take variations (subscript \lq var') of the equilibrium configuration (subscript \lq eq') in the form
\begin{equation}
v(z) = v_{eq}(z) + \epsilon \, v_{var}(z) ,
\end{equation}
subject to the following kinematic boundary conditions at the extremities of the rod,
\beq
\label{eq:resintro}
v_{eq}(0) = 0 , \quad \frac{d v_{eq}(z)}{dz}\bigg|_{z=l} + \chi\,v_{eq}(l) = 0 \quad \text{and} \quad v_{var}(0) = 0 , \quad \frac{d v_{var}(z)}{dz}\bigg|_{z=l} + \chi\,v_{var}(l) = 0 ,
\eeq
where $\chi$ is the signed curvature of the constraint, assumed to be constant. The first variation of the functional \eqref{eq:EPT} is readily obtained as 
\begin{equation}
\label{eq:varprimaEPT}
\delta_{\epsilon}\mathcal{V}(v) = \intop_{0}^{l} \left[B \frac{d^4 v_{eq}(z)}{dz^4} - q \right]v_{var}(z) \, dz + B \frac{d^2 v_{eq}(z)}{dz^2} \frac{d v_{var}(z)}{dz} \bigg|_{z=0}^{z=l} - B \frac{d^3 v_{eq}(z)}{dz^3} v_{var}(z)\bigg|_{z=0}^{z=l} ,
\end{equation}
such that, taking into account the restrictions imposed by Eqs.\,\eqref{eq:resintro}, the vanishing of $\delta_{\epsilon}\mathcal{V}(v)$ for any admissible displacement field $v_{var}(z)$ leads to the differential equation of the linearized elastica (not reported for brevity) and to the non-trivial boundary condition at the right end of the structure, namely,
\begin{equation}
\label{eq:varprimaMomTag}
\frac{d^2 v(z)}{dz^2}\bigg|_{z=l} +  \frac{1}{\chi} \, \frac{d^3 v(z)}{dz^3}\bigg|_{z=l} = 0,	~~ \mbox{or} ~~         M(l) + \frac{1}{\chi} \, T(l) = 0,
\end{equation}
so that the force $T(l)$, tangential to the moving clamp (and coincident now with the shear force transmitted by the elastic rod\footnote{This coincidence is not verified if the clamp is connected to the end of the elastic rod through an elastic hinge.}), is in general not null, but related to the bending moment $M(l)$ through the curvature $\chi$ of the profile along which the clamp may slide. 
The fact that a reaction is present, tangential to a perfectly smooth constraint is an unexpected and noticeable effect sharing similarities with the Eshelby-like force discovered by Bigoni {\it et al.} (2014\,a; 2014\,b; 2015), see also Bosi {\it et al.}(2014). 
This reaction (which is completely unexpected at first glance)
passed unnoticed by Bigoni {\it et al.} (2012) (because they did not experiment a sliding clamp, but only a sliding pin), so that it is the purpose of this article to reconsider the sliding clamp condition, providing for this constraint a full theoretical and experimental validation.

The scope of the present study is: (i.)~to generalize the boundary condition at the curved constraint,  that is Eq.\,\eqref{eq:varprimaMomTag}, showing that it holds true for a profile with variable curvature and when the rod is subject to large displacements, (ii.)~to solve the nonlinear equations of the elastica for a rod with a clamped end movable on a circular constraint, and (iii.)~to experimentally show, through the realization of a proof-of-concept device, that a sliding clamp can be realized in practice to tightly follow the theory. In particular, the presented experiments refer to the case of a rectilinear elastic rod with one clamped end constrained to slide along a bi-circular \lq S-shaped' profile, 
the same system considered by Bigoni {\it et al.} (2012), but from a purely theoretical point of view and with a no-shear assumption at the sliding profile, which is correct only for a movable pin, as shown in the present study. 
The influence of the constraint's curvature is shown both on the critical loads and on the post-critical behaviour, obtained by direct integration of the nonlinear equation of the elastica. An interesting finding is that the postcritical response in compression is \lq almost neutral', in the sense that the load changes very little with increasing displacement of the clamped end, a feature that could be exploited in the realization of a force limiter device, which could for instance be employed in the design of shock absorbers or security belts.

\section{The elastica with a sliding clamp at one end: derivation of boundary conditions}\lb{sec:boundary_conditions}

The boundary condition (\ref{eq:varprimaMomTag}), relating shear force, bending moment and curvature at the sliding constraint can be obtained both with a variational approach and with an asymptotic approach. In fact, the clamp can be replaced by two rollers, close to each other and joined by a rigid bar, in the limit when the length of the bar (and therefore the distance between the two rollers) tends to zero. 
The boundary condition naturally emerges in the former approach, while the latter represents the key for  applications, since it will be experimentally proven that a moving clamp can be realized with two rollers rigidly joined at a small distance from each other.

\subsection{Variational approach}\lb{sec:variational}

Consider an elastic inextensible rod of length $l$ and bending stiffness $B$ that is clamped at its left end, while 
the other end is constrained by another clamp which is free to slide along a frictionless and curved profile parametrically described 
through $\xi$ as $\bx^{c}(\xi)$, see Fig.\,\ref{fig:sketch_elastica} for a sketch of the structure.
%%%%%%%%%% FIG %%%%%%%%%%
\begin{figure}[!htcb]
\renewcommand{\figurename}{\footnotesize{Fig.}}
\begin{center}
\includegraphics[width = 14.5 cm]{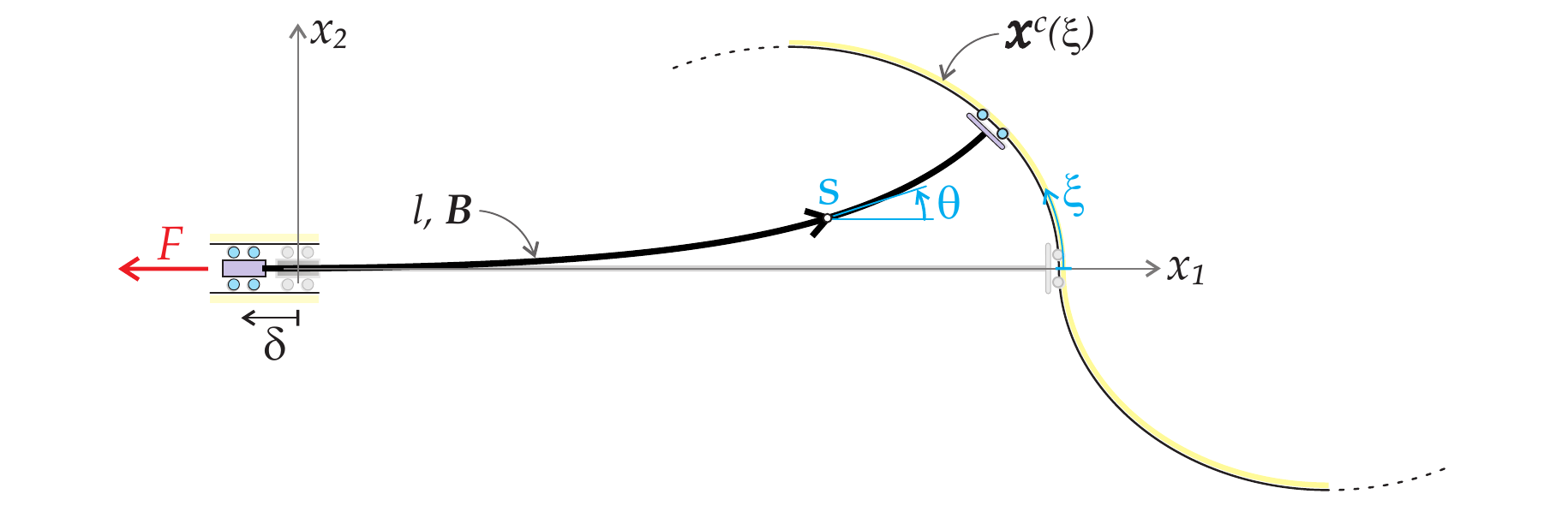}
\caption{\footnotesize 
The planar elastica problem of an inextensible rod of length $l$ and bending stiffness $B$ that is clamped at its left end, and constrained with a clamp at the other end, free of sliding without friction along a curved profile. Note that there is a jump in the curvature of the profile at $\xi =0$, where the tangent to the constraint is vertical.
}
\lb{fig:sketch_elastica}
\end{center}
\end{figure}
%%%%%%%%%% END FIG %%%%%%%%%%

The equations governing the planar elastica problem sketched in Fig.\,\ref{fig:sketch_elastica} can be derived by means of a variational approach.
In particular, the total potential energy $\mathcal{V(\theta,\xi)}$ of the system is
\beq\lb{eq:tpe}
\mathcal{V(\theta,\xi)} = \frac{1}{2}B \intop_{0}^{l} \left(\frac{d\theta(s)}{ds}\right)^2 \, ds - F \left(\intop_{0}^{l} \cos\theta(s) \, ds - x^c_1(\xi) \right) + \lambda \left(\intop_{0}^{l} \sin\theta(s) \, ds - x^c_2(\xi)\right) ,
\eeq
in which $s$ denotes the arc-length of the rod and $\lambda$ is a Lagrange multiplier, enforcing a kinematic compatibility condition at the right end of the system, 
\beq
x_2(l) = \intop_{0}^{l} \sin\theta(s) \, ds = x^c_2(\xi) .
\eeq
Variations (subscript \lq var') from the equilibrium configuration (subscript \lq eq') are considered in the form
\beq\lb{eq:var}
\theta(s,\epsilon) = \theta_{eq}(s) + \epsilon \, \theta_{var}(s) , \quad \xi(\epsilon) = \xi_{eq} + \epsilon \, \xi_{var} ,
\eeq
subject to the following restrictions at the clamp on the left end of the structure
\beq\lb{eq:bc}
\theta_{eq}(0) = 0 , \quad \theta_{var}(0) = 0 .
\eeq

A substitution of Eq.\,\eqref{eq:var} into Eq.\,\eqref{eq:tpe} yields, through integration by parts and taking into account the boundary conditions of Eq.\,\eqref{eq:bc}, the first variation of the
functional $\mathcal{V(\theta,\xi)}$
\begin{flalign}
\lb{eq:fvtpe}
\mathcal{\delta_{\epsilon}V(\theta,\xi)} & = \intop_{0}^{l} \left[-B \, \frac{d^2\theta_{eq}(s)}{ds^2} + F \sin\theta_{eq}(s) + \lambda \cos\theta_{eq}(s)\right]\theta_{var}(s) \, ds \, + \\[1.0mm]
& + B \, \theta_{var}(l) \, \frac{d\theta_{eq}(s)}{ds}\bigg|_{s=l} + \left[F \, \frac{d x^c_1(\xi)}{d\xi}\bigg|_{\xi=\xi_{eq}} - \lambda \, \frac{d x^c_2(\xi)}{d\xi}\bigg|_{\xi=\xi_{eq}} \right] \xi_{var} .
\end{flalign}
Upon noting that the signed curvature of the constraint can be expressed as
\beq
\chi(\xi_{eq}) = -\frac{\theta_{var}(l)}{\xi_{var}},
\eeq
and that
\beq
\lb{chc}
\frac{d x^c_1(\xi)}{d\xi} \bigg |_{\xi=\xi_{eq}} = -\sin \theta_{eq}(l), ~~~ \frac{d x^c_2(\xi)}{d\xi}\bigg|_{\xi=\xi_{eq}} = \cos \theta_{eq}(l),
\eeq
the vanishing of the first variation $\mathcal{\delta_{\epsilon}V(\theta,\xi)}$ for every admissible field $\theta_{var}(s)$ yields the differential equation of the elastica 
(Love, 1927; Audoly and Pomeau, 2010)
\beq
\lb{eq:elastica}
B \, \frac{d^2\theta_{eq}(s)}{ds^2} - F \sin\theta_{eq}(s) - \lambda \cos\theta_{eq}(s) = 0 ,
\eeq
where now the Lagrange multiplier can be identified with the vertical reaction $V$ acting at the left end of the rod (positive when directed upwards along the $x_2$-axis), and an equation expressing the rotational equilibrium of the sliding clamp moving along the curved profile
\beq
\lb{quellabuona}
B \, \frac{d\theta_{eq}(s)}{ds}\bigg|_{s=l} - \frac{1}{\chi(\xi_{eq})} \left[F \, \frac{d x^c_1(\xi)}{d\xi} \bigg |_{\xi=\xi_{eq}} - V  \frac{d x^c_2(\xi)}{d\xi}\bigg|_{\xi=\xi_{eq}} \right] = 0 ,
\eeq
which, using Eqs. (\ref{chc}), becomes
\beq\lb{eq:eqclamp2}
B \, \frac{d\theta_{eq}(s)}{ds}\bigg|_{s=l} + \frac{1}{\chi(\xi_{eq})} \left[F \, \sin \theta_{eq}(l) + V \cos \theta_{eq}(l) \right] = 0 ,
\eeq
a condition showing that the bending moment and the shear force at the moving clamp relates as through Eq.\,\eqref{eq:varprimaMomTag}, but now $\chi(\xi_{eq})$ is the curvature of the profile where the clamp is located.

\subsection{Asymptotic approach}\lb{sec:asymptotic}

\begin{figure}[!htcb]
\renewcommand{\figurename}{\footnotesize{Fig.}}
\begin{center}
\includegraphics[width=10.0 cm]{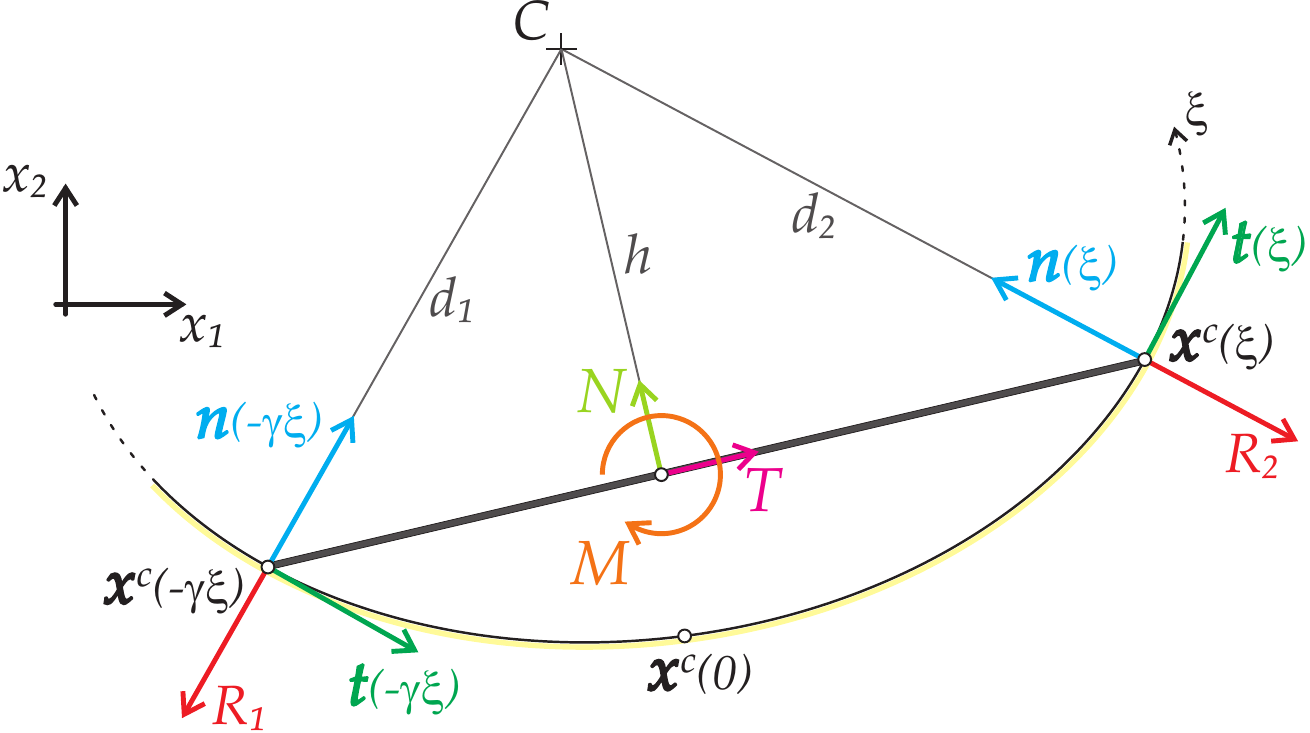}
\caption{\footnotesize An elastic rod transmits the internal actions $N$, $T$, and $M$ to a rigid bar connecting two rollers that are subject to the reactions $R_1$ and $R_2$ from a rigid and perfectly smooth profile of co-ordinates $\bx^c(\xi)$.}
\label{F_CONSTRAINT}
\end{center}
\end{figure}

Consider a generic profile of co-ordinates $\bx^{c}(\xi)$, parametrized by the arc-length $\xi$, along which two rollers, joined by a rigid bar, may freely slide without friction. It is assumed that an axial force $N$, a shear force $T$, and a bending moment $M$ are applied on the rigid bar (corresponding to the internal actions transmitted by the elastic rod), together with the two reactions transmitted by the  rollers, $R_1$ and $R_2$, acting along the normal $\bf n$ to the profile, see Fig.\,\ref{F_CONSTRAINT}.
%%%%%%%%%% FIG %%%%%%%%%%
%%%%%%%%%% END FIG %%%%%%%%%%

Initially, the rigid bar is considered of finite length, but eventually this length will be shrunk to zero. This rigid bar is characterized by the vector $\hat{\bx}(\xi,\gamma) = \bx^c(\xi)-\bx^c(-\gamma \, \xi)$, in which $\gamma$ is a positive scalar, so that its equilibrium  requires 
\beq
\lb{E_EQUILIBRIUM}
\left\{
\barr{l}
\ds R_{1} \frac{\hat{\bx}(\xi,\gamma)}{|\hat{\bx}(\xi,\gamma)|} \scalp \bt(-\gamma \, \xi) + R_{2} \frac{\hat{\bx}(\xi,\gamma)}{|\hat{\bx}(\xi,\gamma)|} \scalp \bt(\xi) = N , \\[6.0mm]
\ds R_{1} \frac{\hat{\bx}(\xi,\gamma)}{|\hat{\bx}(\xi,\gamma)|} \scalp \bn(-\gamma \, \xi) + R_{2} \frac{\hat{\bx}(\xi,\gamma)}{|\hat{\bx}(\xi,\gamma)|} \scalp \bn(\xi) = T , \\[6.0mm]
\ds h = \frac{M}{T} ,
\earr
\right.
\eeq
where the unit vector tangent to the profile is denoted by $\bf t$, whereas $h$ is the distance between the intersection point of the two normals, {\it C\,}, and the rigid bar. Notice that $|\hat{\bx}(\xi,\gamma)|$ corresponds to the length of the bar, while solution of Eq.\,\eqref{E_EQUILIBRIUM} immediately yields explicit formulae for the two reactions $R_1$ and $R_2$ between the rollers and the curved profile
\beq
\lb{E_REACTIONS}
\left\{
\barr{l}
\ds R_{1} = \frac{|\hat{\bx}(\xi,\gamma)|[N\,\hat{\bx}(\xi,\gamma) \scalp \bn(\xi) - T\,\hat{\bx}(\xi,\gamma) \scalp \bt(\xi)]}{[\hat{\bx}(\xi,\gamma) \scalp \bn(\xi)][\hat{\bx}(\xi,\gamma) \scalp \bt(-\gamma \, \xi)] - [\hat{\bx}(\xi,\gamma) \scalp \bn(-\gamma \, \xi)][\hat{\bx}(\xi,\gamma) \scalp \bt(\xi)]} , \\[6.0mm]
\ds R_{2} = \frac{|\hat{\bx}(\xi,\gamma)|[N\,\hat{\bx}(\xi,\gamma) \scalp \bn(-\gamma \, \xi) - T\,\hat{\bx}(\xi,\gamma) \scalp \bt(-\gamma \, \xi)]}{[\hat{\bx}(\xi,\gamma) \scalp \bn(-\gamma \, \xi)][\hat{\bx}(\xi,\gamma) \scalp \bt(\xi)] - [\hat{\bx}(\xi,\gamma) \scalp \bn(\xi)][\hat{\bx}(\xi,\gamma) \scalp \bt(-\gamma \, \xi)]} .
\earr
\right.
\eeq

To proceed, we notice from Fig.\,\ref{F_CONSTRAINT} that $\bx^c(-\gamma \, \xi) + d_{1}\,\bn(-\gamma \, \xi) = \bx^c(\xi) + d_{2}\,\bn(\xi)$, such that, taking the scalar product with $\bt(\xi)$, an expression is obtained for the distance $d_1$, namely
\beq
\ds d_{1} = \frac{\hat{\bx}(\xi,\gamma) \scalp \bt(\xi)}{\bn(-\gamma \, \xi) \scalp \bt(\xi)}, 
\eeq
and, in turn, for the distance $h$ that relates bending moment $M$ and shear force $T$,
\beq
\lb{E_H}
h =  d_1 \, \frac{\hat{\bx}(\xi,\gamma)}{|\hat{\bx}(\xi,\gamma)|} \scalp \bt(-\gamma \, \xi) .
\eeq

Finally, in the limit of vanishing distance $|\hat{\bx}(\xi,\gamma)|$ the two reactions $R_1$ and $R_2$ blow up to infinity, whereas the distance $h$ converges to
\beq
\lb{E_H_LIN}
\ds \lim_{\xi \to 0} h = \frac{1}{|\bt'(0)|} ,
\eeq
where a prime denotes differentiation with respect to $\xi$. Noting that $|\bt'(0)|$ corresponds to the curvature of the profile evaluated at $\xi=0$, we recover with the asymptotic analysis Eq.\,\eqref{eq:varprimaMomTag} and Eq.\,\eqref{eq:eqclamp2} relating bending moment and shear force.

In an experimental setting, a sliding clamp can be realized with two rollers joined with a short and rigid bar. In this case, the distance $h$ may be calculated through Eq.\,(\ref{E_H}), so that the condition  (\ref{E_EQUILIBRIUM})$_3$ provides an estimate of the real boundary condition obtained in the experiment.
For the experimental tests 
that will be described later, two rollers were joined at a distance of 30\,mm and these slide on a circle of radius 150\,mm. Therefore, the real boundary condition is $M/T =$ 149.2\,mm, to be compared with the ideal boundary condition of $M/T =$ 150\,mm, showing 
that the behaviour of the physical model of the clamp follows very closely the mathematical model.

\section{Bifurcation and post-critical behaviour of an elastic rod with a sliding clamp}\lb{S_BUCK_POST}

An inextensible elastic rod, rectilinear in the undeformed configuration, has a moving clamp on its right end, constrained to slide without friction along an \lq S-shaped' bi-circular profile, Fig.\,\ref{fig:sketch_elastica_s}. 
%%%%%%%%%% FIG %%%%%%%%%%
\begin{figure}[!htcb]
\renewcommand{\figurename}{\footnotesize{Fig.}}
    \begin{center}
  \includegraphics[width = 13.5 cm]{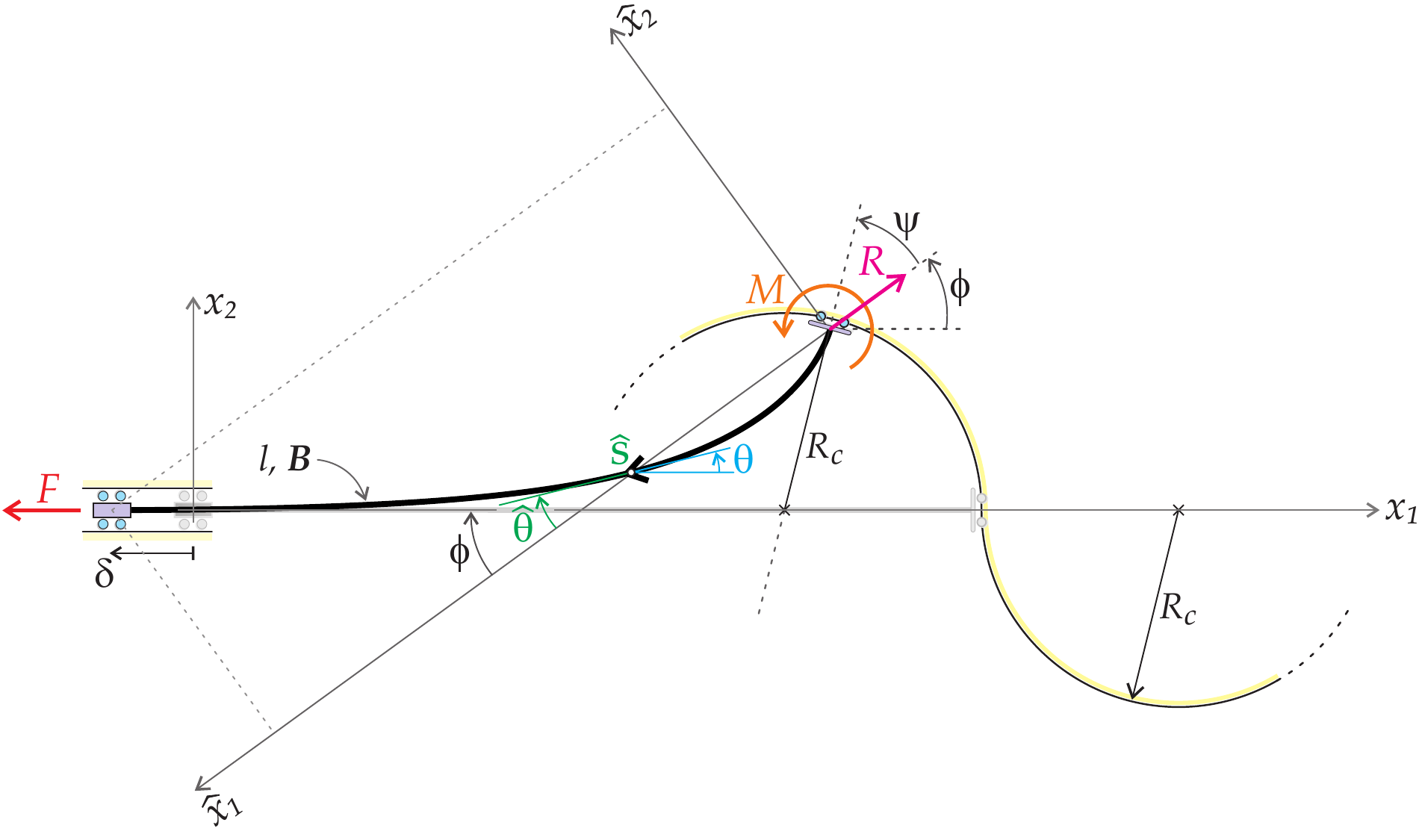}
    \caption{\footnotesize Bifurcation and post-bifurcation problem for an inextensible elastic rod with two moving clamps at the ends: one can slide horizontally at the left end, while the other slides along a frictionless, bi-circular profile. The trivial equilibrium configuration is sketched in gray.}
    \lb{fig:sketch_elastica_s}
    \end{center}
\end{figure}
%%%%%%%%%% END FIG %%%%%%%%%%
The straight configuration is the trivial equilibrium solution, in which the structure is subject to an axial tensile (positive) or compressive (negative) dead force $F$. 
The buckling and the postcritical behavior of the structure is examined in detail below. 

\subsection{Bifurcation loads}\lb{SS_BUCK}

Denoting by $v(z)$ the transverse displacement, the linearized differential equation governing equilibrium of an elastic rod subject to an axial force $F$ is
\beq
\label{diff_equilibrium_eq}
\frac{d^4 v(z)}{dz^4} - \alpha^2\,\mbox{sgn}(F)\,\frac{d^2 v(z)}{dz^2} = 0 ,
\eeq
where $\alpha^2 = |F|/B$ and \lq sgn' indicates the sign function.
The boundary conditions at the two ends of the rod read
\beq
\lb{bc_d_i}
v(0) = 0 , ~~~
\frac{d v(z)}{dz}\bigg|_{z=0} = 0 , ~~~ 
-\frac{d^3 v(z)}{dz^3}\bigg|_{z=l} = \chi \ds \frac{d^2 v(z)}{dz^2}\bigg|_{z=l}, ~~~ 
\frac{d v(z)}{dz}\bigg|_{z=l} =-\chi \, v(l),
\eeq
involving the signed curvature $\chi=\pm 1/R_c$ of the circle.
Note that Bigoni {\it et al.} (2012) have referred to a rotational spring of stiffness $k_r$ at the right end of the rod, but they did not consider a shear force acting on that end. 
Consideration of the elastic hinge connecting the rod to the moving clamp implies that (\ref{eq:varprimaMomTag})$_1$ and (\ref{eq:eqclamp2}) do not hold, although (\ref{eq:varprimaMomTag})$_2$ and (\ref{quellabuona}) still continue to hold, so that the Eq.\,(15)$_2$ of Bigoni {\it et al.} (2012) has to be modified by multiplying the term on the right hand side by $[-1 - \mbox{sgn}(F) k_r \widehat{\chi}/(B l\alpha^2)]$. 

A substitution of the general solution of Eq.\,(\ref{diff_equilibrium_eq}) into the boundary conditions (\ref{bc_d_i}) yields the condition for the critical loads\footnote{In the case of an elastic hinge connecting the rod to the moving clamp considered by Bigoni {\it et al.} (2012), their Eq.\,(17) has to be modified by replacing the quantity multiplying the rotational spring stiffness $k_r$  with the term at the left hand side of Eq.\,\eqref{gen_Sol} multiplied by $\widehat{|\chi|} \mbox{sgn}(F) /(B l \alpha^2)$.}, that is
\beqar
\barr{ll}
\lb{gen_Sol}
\alpha l\,\mbox{sgn}(F)\cosh(\sqrt{\mbox{sgn}(F)}\,\alpha l)+\sqrt{\mbox{sgn}(F)}\big[-1+\mbox{sgn}(F)(1+\widehat{\chi})\ds \frac{\alpha^2 l^2}{\widehat{\chi}^2 } \big]\sinh(\sqrt{\mbox{sgn}(F)}\,\alpha l)=0 .
\earr
\eeqar

\begin{figure}[!htcb]
\renewcommand{\figurename}{\footnotesize{Fig.}}
    \begin{center}
   \includegraphics[width = 13 cm]{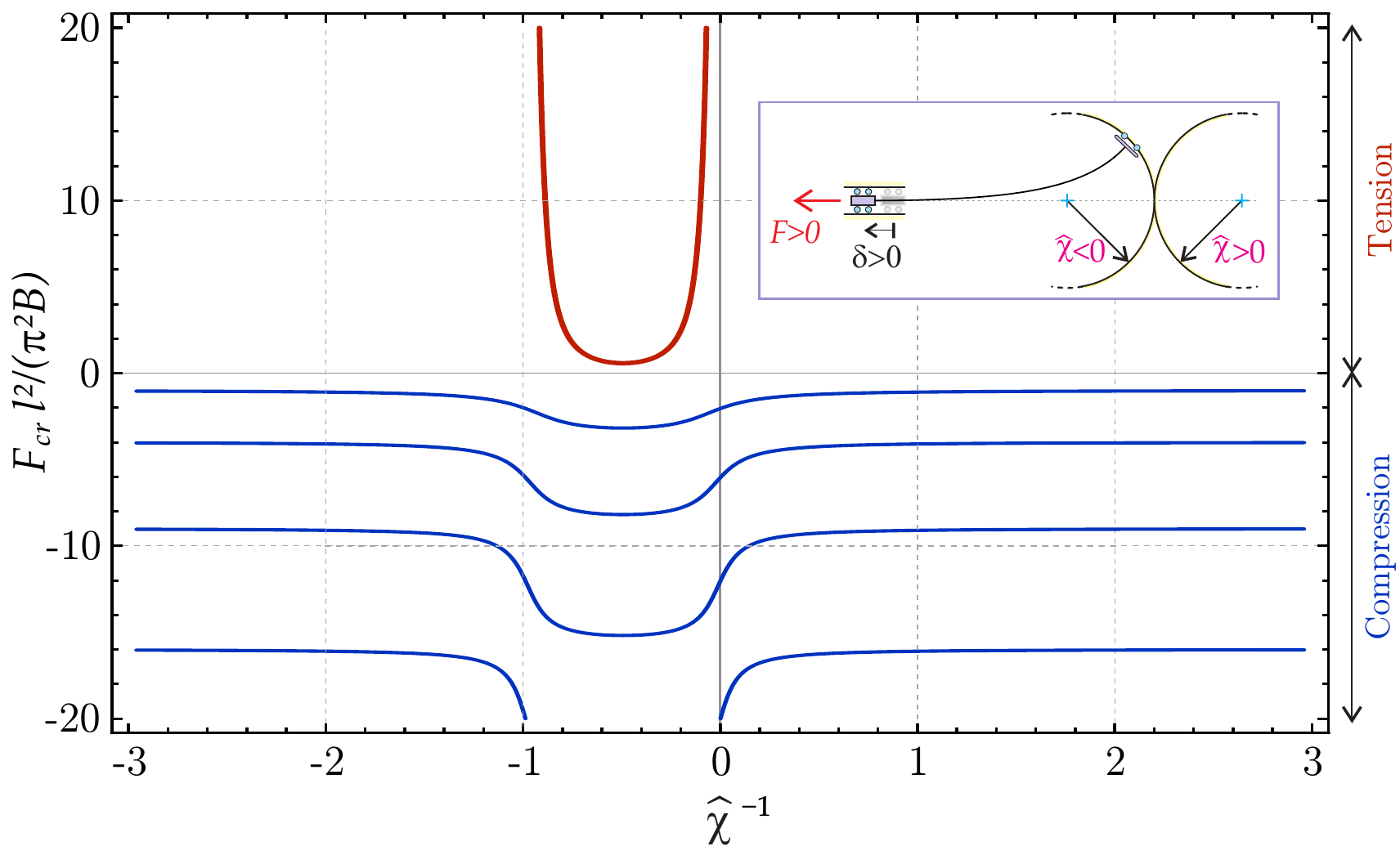}
    \caption{\footnotesize 
				Dimensionless bifurcation load $F_{\rm{cr}}$ (a negative sign denotes compression), reported for the first five modes, of the structure sketched in the inset as a function of the signed dimensionless curvature $\widehat{\chi}$ of the circular profile.}
    \label{F_BUCK}
    \end{center}
\end{figure}

Buckling loads, made dimensionless through multiplication by $l^2/(\pi^2 B)$, and effective length factors $l_0/l = \pi/l\sqrt{B/F_{cr}}$ are reported in Fig.\,\ref{F_BUCK} and in Tables\,\ref{crit_loads1} and  \ref{crit_loads2} as functions of the dimensionless signed curvature of the constraint $\widehat{\chi} = l \chi$. 
Note that this figure and the tables replace data reported by Bigoni {\it et al.} (2012) in their Fig.\,9 (right) and Table\,2, which refer to a zero-shear force assumption, which has been proved to be 
incorrect in the present article. The correct results reported in Fig. \ref{F_BUCK} show a symmetry of the tensile buckling loads about $\widehat{\chi} = 2$, which was absent with the incorrect assumption of zero-shear force.

%%%%%%%%%% FIG %%%%%%%%%%
%%%%%%%%%% END FIG %%%%%%%%%%

\begin{table} [!htcb]
\renewcommand{\tablename}{\footnotesize{Table}}
\begin{center}
\begin{tabu}{c|cccccccccccccc}
\toprule[.8pt]
$\widehat{\chi}^{-1}=-R_c/l$ & -$\infty$ & -2 & -1& -0.75 & -0.5 & -0.25& \small{negative curvature} \\
\hline
\rowfont{\color{traz}}
\textcolor{black}{\multirow{2}{*}{$\ds \frac{l_0}{l}$}}  & - & -  & - & 0.784 & 1.309 & 0.784 & \small {tensile buckling} \\
 \rowfont{\color{comp}}            & 1 &  0.956 & 0.699 & 0.581 & 0.561 & 0.581 &\small {compressive buckling} \\
\hline
\rowfont{\color{traz}}
\textcolor{black}{\multirow{2}{*}{$\ds \frac{F_{cr}l^2}{\pi^2 B}$ }}& - & - & + $\infty$& +1.625 & +0.583 & +1.625 & \small {tensile buckling} \\
   \rowfont{\color{comp}}                 & -1 &  -1.094 & -2.046 & -2.959 &-3.174 & -2.959 &\small {compressive buckling}\\
\bottomrule
\end{tabu}
\caption{\footnotesize Dimensionless buckling loads $F_{cr}l^2/(\pi^2 B)$ and effective length factors $l_0/l$ for negative curvature.}
\label{crit_loads1}
\end{center}
\end{table}

\begin{table} [!htcb]
\renewcommand{\tablename}{\footnotesize{Table}}
\begin{center}
\begin{tabu}{c|cccccccccccc}
\toprule[.8pt]
$\widehat{\chi}^{-1}=R_c/l$  &+0.25 & +0.5 & +0.75 & +1  & +2 & +$\infty$ & \small{positive curvature}\\
\hline
\rowfont{\color{traz}}
\textcolor{black}{\multirow{2}{*}{$\ds \frac{l_0}{l}$}}  & - &-  & - & - &-  &  - & \small {tensile buckling}\\
 \rowfont{\color{comp}}         & 0.838 & 0.904 & 0.937 & 0.956  & 0.984 & 1 &\small {compressive buckling} \\
\hline
\rowfont{\color{traz}}
\textcolor{black}{\multirow{2}{*}{$\ds \frac{F_{cr}l^2}{\pi^2 B}$}}& -  & - & -& - & - & - & \small {tensile buckling} \\
   \rowfont{\color{comp}}                &-1.424 & -1.223 & -1.138 & -1.094 & -1.033 & -1 &\small {compressive buckling}\\
\bottomrule
\end{tabu}
\caption{\footnotesize Dimensionless buckling loads $F_{cr}l^2/(\pi^2 B)$ and effective length factors $l_0/l$ for positive curvature.}
\label{crit_loads2}
\end{center}
\end{table}
From the reported results, we can note that tensile buckling is always excluded for positive curvature, while this becomes possible for negative.

\subsection{The elastica for a \lq S-shaped' bi-circular profile}\lb{SS_ELASTICA}

Write $R$ as the resultant force acting on the rod and, in turn, write $\phi$ as its inclination with respect to the horizontal axis $x_1$, see Fig.\,\ref{fig:sketch_elastica_s}. For the solution of the problem, it is now expedient to introduce the following change of variables, namely $s = l - \hat{s}$ and $\theta(s) = \phi - \hat{\theta}(\hat{s})$, such that the non-linear equation of the elastica \eqref{eq:elastica} now becomes
\beq\lb{eq:elastica_girata}
\frac{d^2\hat\theta(\hat s)}{d\hat{s}^2} - \frac{R}{B} \sin\hat\theta(\hat s) = 0 ,
\eeq
where for simplicity the subscript \lq {\it eq\,}' has been dropped. Notice that Eq.\,\eqref{eq:elastica_girata} corresponds to the elastica of a inextensible rod written in a local reference system $(\hat{x}_1,\hat{x}_2)$, such that the angle $\hat \theta(\hat s)$ measures the rotation of the normal to the rod with respect to the axis $\hat{x}_1$ at a distance $\hat s$ from the sliding clamp, see Fig.\,\ref{fig:sketch_elastica_s}, in which the local reference system has also been reported. The elastic line problem can now be solved upon noting that:

\begin{enumerate}[i.)]

\item a condition of kinematic compatibility can be obtained by observing from Fig.\,\ref{fig:sketch_elastica_s} that the coordinates of the elastica evaluated at $\hat s = l$, namely, $\hat x_{1}(l)$ and $\hat x_{2}(l)$, are related to the two angles $\phi$ and $\psi$ reported in the figure, and to the radius $R_c$ of the constraint via
\beq
\lb{eq:comp}
\left[\hat x_{1}(l) \mp R_{c} \cos\psi \right]\tan\phi \mp R_{c}\sin\psi - \hat x_{2}(l) = 0,
\eeq
where $\phi$ and $\psi$ are assumed positive as shown in the figure. Note that in Eq.\,\eqref{eq:comp} the upper (lower) sign holds for the case of the sliding clamp lying on the left (right) half-circle;

\item the curved constraint transmits to the rod a bending moment $M$ and a force $R$, parallel to $\hat x_{1}$ and assumed positive when opposite to the direction of that axis, so that for $0 \le \phi < \pi/2$ ($\pi/2 < \phi \le \pi$) this corresponds to a positive tensile (negative compressive) dead force $F$ applied to the structure defined by
\beq
\lb{eq:fr}
F = R\cos\phi;
\eeq

\item with the symbols introduced in Fig.\,\ref{fig:sketch_elastica_s}, the following condition holds between the angle $\hat \theta$ evaluated at $\hat s = l$ and the angle $\phi$
\beq
\lb{eq:bc_l}
\hat \theta(l) = \phi ,
\eeq
and similarly, the following condition holds between the angle $\hat \theta$ evaluated at $\hat s = 0$ and the angle $\psi$
\beq
\lb{eq:bc_0}
\hat \theta(0) = -\psi ;
\eeq

\item as demonstrated in Section~\ref{sec:boundary_conditions}, equilibrium of the constraint sliding on the circular profile of radius $R_{c}$ requires that
\beq
\lb{eq:cond_constraint}
M = R_{c} \, |R| \sin\psi,
\eeq
so that Eqs.\,\eqref{eq:bc_0}-\eqref{eq:cond_constraint} immediately provide the boundary condition
\beq
\lb{eq:bc_0_t}
\hat \theta(0) = -\arcsin\! \left(\frac{B}{R_c \, |R|}\frac{d\hat\theta(\hat s)}{d \hat s}\bigg|_{\hat s = 0}\right) \!.
\eeq

\end{enumerate}

\noindent
Integration of Eq.\,\eqref{eq:elastica_girata} from $0$ to $\hat s$, after multiplication by $d \hat \theta(\hat s) / d \hat s$, leads to
\beq
\lb{eq:first_int}
\left(\frac{d \hat\theta(\hat s)}{d \hat s}\right)^2 = 2\,\tilde{\alpha}^2\left[\frac{2}{k^2} - 1 - \mbox{sgn}(R)\cos \hat\theta(\hat s) \right] ,
\eeq
where $\tilde{\alpha}^2 = |R| / B$
and
\beq
\lb{eq:k}
k^2 = \ds \frac{4\tilde{\alpha}^2}{\left(R_{c}\, |R| \sin\psi / B \right)^2 + 2\tilde{\alpha}^2\left[\mbox{sgn}(R)\cos\hat\theta(0) + 1 \right]} ,
\eeq
in which Eq.\,\eqref{eq:cond_constraint} has been used to express the signed curvature of the rod at $\hat s = 0$.
The introduction of the change of variable
\beq
\beta(\hat s) = [\hat \theta(\hat s) - \mbox{H}(R)\,\pi]/2,
\eeq
where $\mbox{H}$ denotes the Heaviside step function, allows to re-write Eq.\,\eqref{eq:first_int} as
\beq
\left(\frac{d\beta(\hat s)}{d \hat s}\right)^2 = \ds \frac{\tilde{\alpha}^2}{k^2} \left(1 - k^2\sin^2\beta(\hat s)\right),
\eeq
such that a second change of variable $u = \hat s \, \tilde{\alpha} / k$ yields
\beq
\lb{eq:dbe_du}
\frac{d\beta(u)}{du} = \ds \pm \sqrt{1 - k^2\sin^2\beta(u)} \, .
\eeq

Restricting for conciseness the treatment to the case \lq$+$', Eq.\,\eqref{eq:dbe_du} provides the following solution for $\beta(u)$,
\beq
\lb{eq:be}
\beta(u) = \ds \mbox{am}\left[u + \mbox{F}\left[\beta(0), k\right], k\right],
\eeq
where am and F are the Jacobi elliptic function amplitude and the incomplete elliptic integral of the first kind of modulus $k$, respectively (Byrd and Friedman, 1971). Keeping now into account that $d \hat x_{1}/d \hat s = \cos \hat\theta(\hat s)$ and that $d \hat x_{2}/d\hat s = \sin \hat\theta(\hat s)$, an integration provides the two coordinates $\hat x_{1}$ and $\hat x_{2}$ of the elastica expressed in terms of the arc-length $u$, namely
\beq
\lb{eq:coords}
\left\{
\barr{l}
\ds \hat x_{1}(u) = \mbox{sgn}(R)\,\frac{2}{k\tilde{\alpha}} \{(1 - k^2/2)u + \mbox{E}\left[\beta(0), k\right] - \mbox{E}\left[\mbox{am}\left[u + \mbox{F}\left[\beta(0), k\right], k\right], k\right]\} , \\[6mm]
\ds \hat x_{2}(u) = \mbox{sgn}(R)\,\frac{2}{k\tilde{\alpha}} \{\mbox{dn}\left[u + \mbox{F}\left[\beta(0), k\right], k\right] - \mbox{dn}\left[\mbox{F}\left[\beta(0), k\right], k\right]\} ,
\earr
\right.
\eeq
in which the constants of integration are chosen so that $\hat x_{1}$ and $\hat x_{2}$ vanish at $\hat s = 0$. In Eqs.\,\eqref{eq:coords} dn is the Jacobi elliptic function delta-amplitude of modulus $k$, while E is the incomplete elliptic integral of the second kind (Byrd and Friedman, 1971).

The horizontal displacement $\delta$ of the clamp on the left-hand side of the structure (assumed positive for a lengthening of the system) is given in the form
\beq
\lb{eq:delta}
\delta = \frac{\hat x_1(l) \mp R_{c}\cos\psi}{\cos\phi} \pm R_{c} - l,
\eeq
where, similarly to Eq.\,\eqref{eq:comp}, the upper (lower) sign holds for the case of the sliding clamp lying on the left (right) half-circle.

On the basis of the equations reported above, the axial dead load $F$ can be computed as a function of the end displacement $\delta$ through the following steps:

\begin{enumerate}[i.)]

\item a value is fixed for the signed curvature of the rod at the sliding clamp, i.e. $d \hat \theta(\hat s)/d\hat s |_{\hat s = 0}$, such that the modulus $k$ can be expressed using Eq.\,\eqref{eq:bc_0_t} as a function of the unknown $R$;

\item consequently, the expressions \eqref{eq:coords} for the coordinates of the elastica and Eq.\,\eqref{eq:be} become functions of $R$ only, when evaluated at $\hat s = l$;

\item the angle $\phi$ is provided by Eq.\,\eqref{eq:bc_l}, so that the kinematic compatibility condition \eqref{eq:comp} becomes a nonlinear equation in the variable $R$, which can be numerically
solved (we have used the function FindRoot of $\mbox{Mathematica}^{\mbox{\tiny\textregistered}}$\,6.0);

\item once $R$ is numerically determined, the external dead load $F$ and the displacement $\delta$ can be obtained from Eq.\,\eqref{eq:fr} and Eq.\,\eqref{eq:delta}, respectively.

\end{enumerate}

An example of integration of the elastica is reported together with experimental results in the next Section, see Fig.\,\ref{fig:post_clamp} for a detailed comparison. In conclusion, notice that the present treatment remains almost unchanged when the sliding clamp is replaced with a pin equipped with a rotational spring of stiffness $k_r$.
This constraint was analyzed by Bigoni {\it et al.} (2012) under the incorrect zero-shear assumption and simply requires the replacement of the boundary condition \eqref{eq:bc_0} with
\beq
\lb{eq:bc_0_kr}
\hat \theta(0) = \frac{M}{k_r} - \psi ,
\eeq
such that in this case Eq.\,\eqref{eq:bc_0_t} becomes
\beq
\lb{eq:bc_0_t_kr}
\hat \theta(0) = \frac{M}{k_r} - \arcsin\!\left(\frac{B}{R_c \, |R|}\frac{d\hat\theta(\hat s)}{d \hat s}\bigg|_{\hat s = 0}\right) \!.
\eeq

\section{The realization and testing of the proof-of-concept compliant mechanism}\lb{S_EXPERIMENTS}

The compliant mechanism sketched in Fig.\,\ref{fig:sketch_elastica_s} 
was designed and constructed as follows. 
A sliding clamp was realized with two roller bearings placed at a distance of 30\,mm to each other, 
free to slide along a bi-circular slot in a 2mm steel plate.
%%%%%%%%%% FIG %%%%%%%%%%
\begin{figure}[!htcb]
\renewcommand{\figurename}{\footnotesize{Fig.}}
    \begin{center}
    \includegraphics[width = 10 cm]{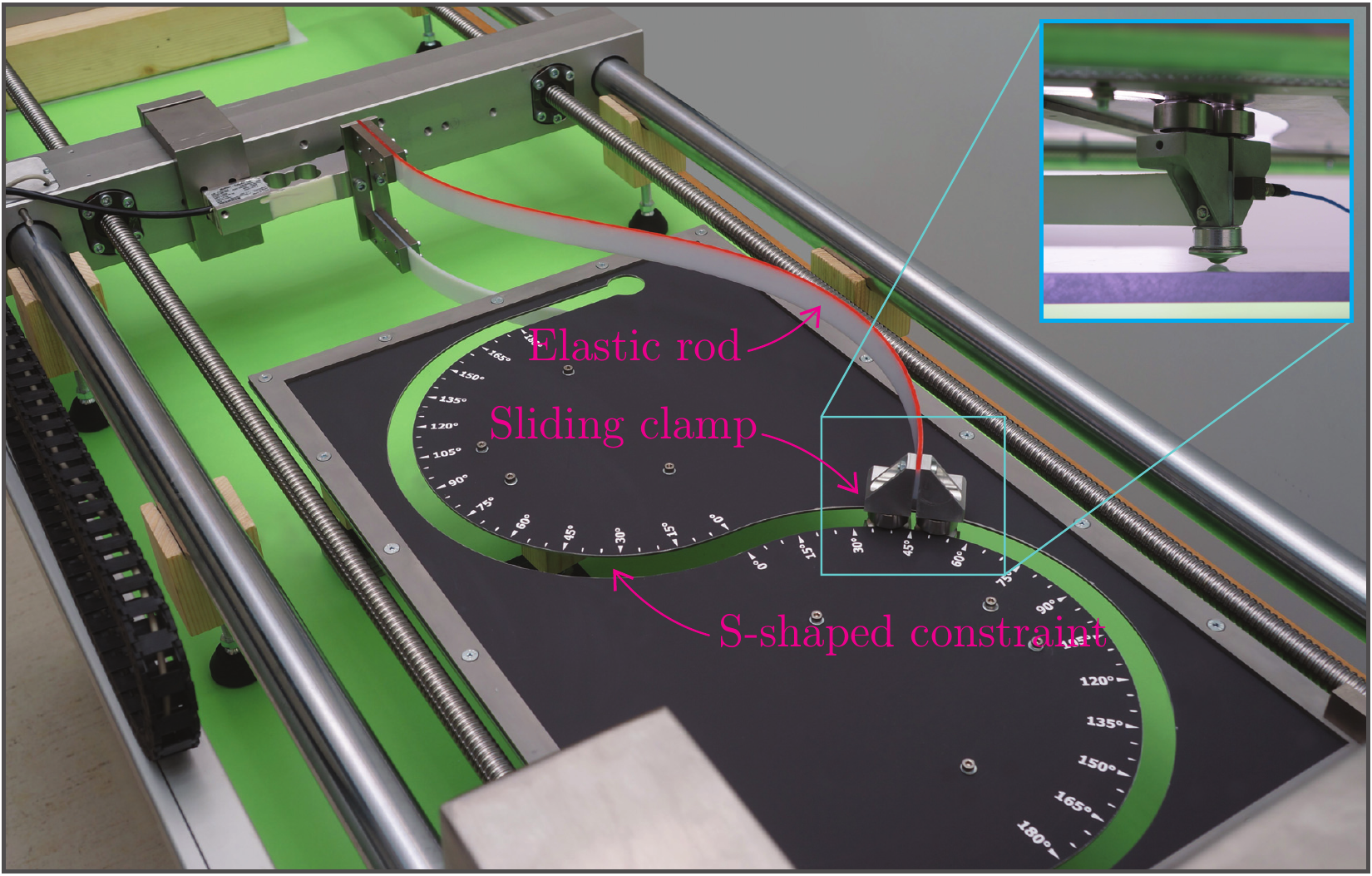}
    \caption{\footnotesize A side view of the experimental set-up for the buckling and post-buckling behaviour of an elastic rod with a movable clamp on its right end, see the model sketched in Fig.\,\ref{fig:sketch_elastica_s}.}
    \lb{expset}
    \end{center}
\end{figure}
%%%%%%%%%% END FIG %%%%%%%%%%
The elastic rod was made up of two parallel, 600\,mm long polycarbonate strips with a rectangular cross section of 3\,mm $\times$ 25\,mm each, for an overall bending stiffness $B \simeq$ 265\,kN\,mm$^2$. The structure was loaded at a speed of 1\,mm/s by imposing at one end a horizontal displacement using a Midi 10 electromechanical testing machine (10\,kN maximum force, from Messphysik Materials Testing). In order to prevent the weight of the structure 
influencing its static properties, the testing machine was turned into a horizontal position and the movable 
clamp was supported with an almost frictionless bearing, see the inset of Fig.\,\ref{expset}.  
Loads and displacements were measured with a MT\,104 load cell (0.5\,kN maximum load, from Mettler-Toledo) and a potentiometric displacement transducer LTM-900-S\,IP65 (from Gefran).
Furthermore, an IEPE\,333B50 accelerometer (from PCB Piezotronics Inc.) was attached at one end of the structure to both monitor the vibrations during the test (which should remain sufficiently small)
and precisely detect the instant of buckling. Data were acquired with a NI CompactDAQ system interfaced with Labview 8.5.1 (National Instruments).
The experiments were performed at the \lq Instabilities Lab' of the University of Trento. Fig.\,\ref{expset} shows a photo of the experimental setup, whereas additional material with movies of the experiments 
is provided in the electronic supplementary material and can be found at http://ssmg.unitn.it/.
%%%%%%%%%% FIG %%%%%%%%%%
\begin{figure}[!htcb]
\renewcommand{\figurename}{\footnotesize{Fig.}}
\begin{center}
 \includegraphics[width = 13 cm]{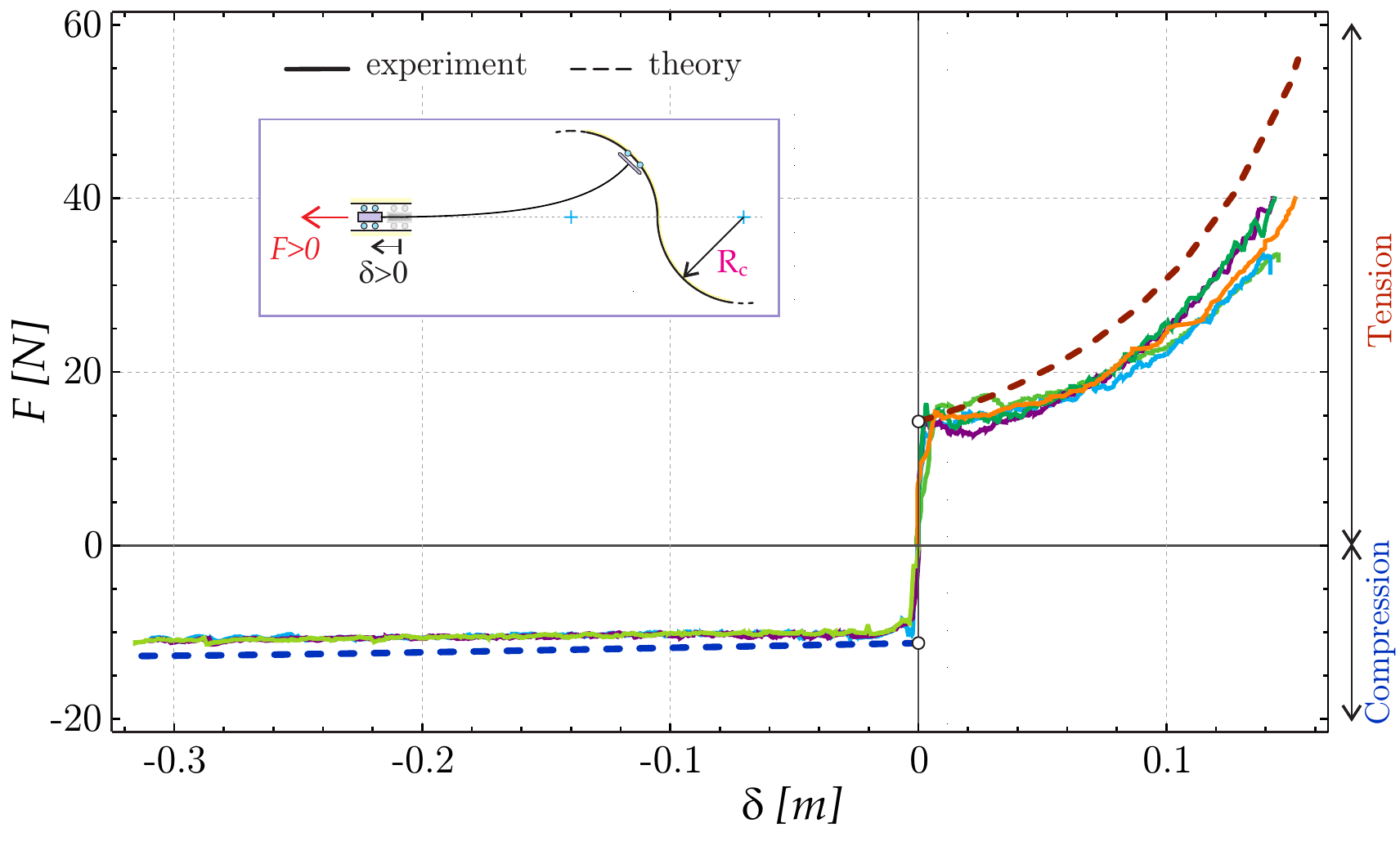}
\caption{\footnotesize The postcritical behaviour of the compliant mechanism sketched in Fig.\,\ref{fig:sketch_elastica_s}, corresponding to the first mode under tensile and compressive loads. 
In particular, the applied axial load $F$ is shown versus the end displacement of the system $\delta$. Note the flat, \lq almost neutral' postcritical response in compression.
The theoretical solution (reported dashed) has been calculated with Eq. (\ref{eq:comp}), through steps (i.)-(iv.).
}
\lb{fig:post_clamp}
\end{center}
\end{figure}
%%%%%%%%%% END FIG %%%%%%%%%%

Results of tension/compression experiments are reported in Figs.\,\ref{fig:post_clamp}, \ref{deformate_comp}, and \ref{sovrapp}. In particular, the applied load $F$ versus end displacement $\delta$ relation is shown in Fig.\,\ref{fig:post_clamp}. Here, beside the remarkable agreement with the theoretical results (reported with a dashed line), we may note that the structure stiffens during the postcritical behaviour in tension, but displays an \lq almost neutral' behaviour in compression. With \lq almost neutral' it is meant that the external load increases very weakly during compression, so that it remains almost constant while the displacement increases. For this reason, the mechanical system is very stiff up to the buckling load in compression, but the load does not vary after this even if the displacement progresses by a large amount. We can therefore notice that the compliant system operates as a force limiter. 
%%%%%%%%%% FIG %%%%%%%%%%
\begin{figure}[!htcb]
\renewcommand{\figurename}{\footnotesize{Fig.}}
    \begin{center}
    \includegraphics[angle= 0, width = 15.5 cm]{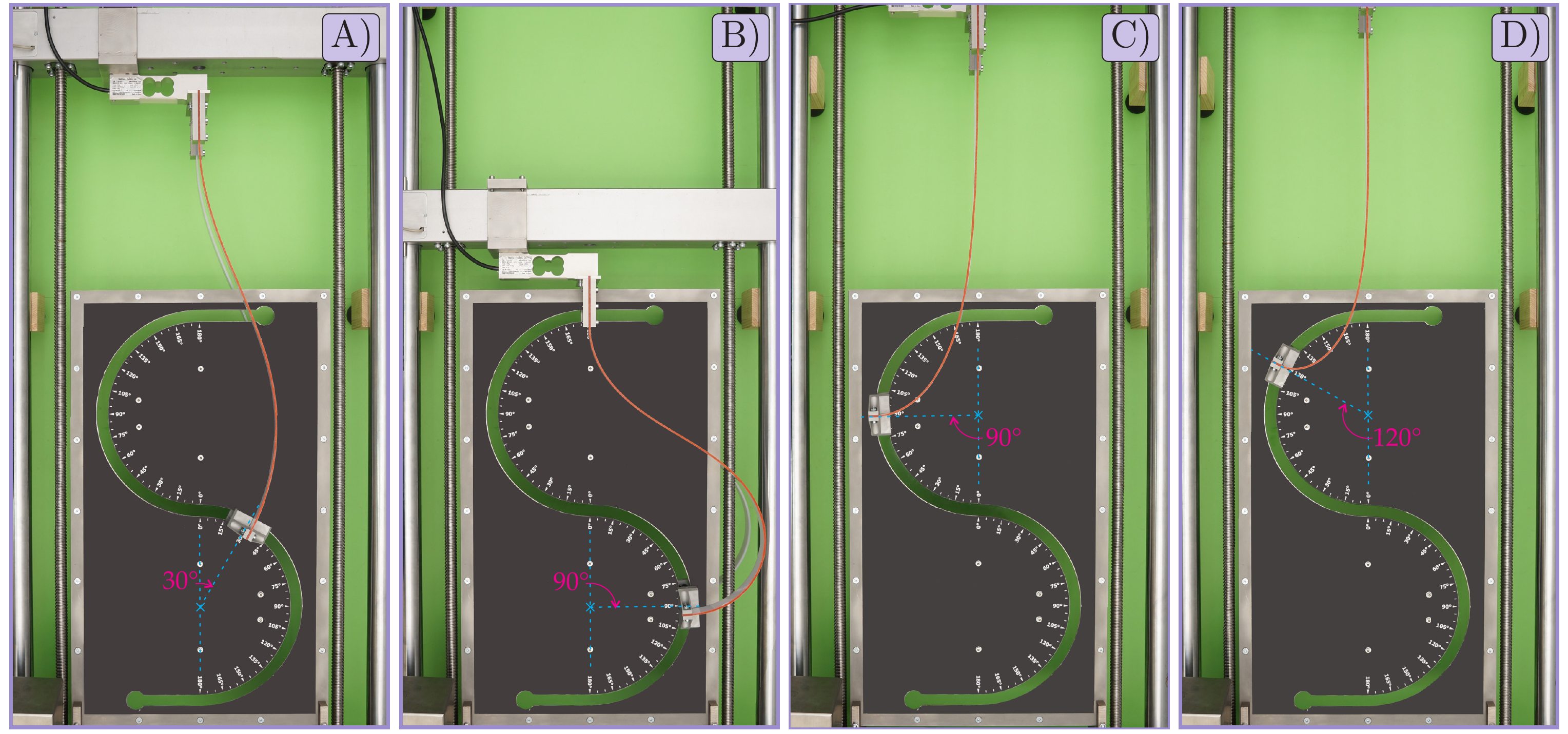}
    \caption{\footnotesize The deformed elastica during a compression (A and B) and a tension (C and D) test performed on the structure sketched in Fig.\,\ref{fig:sketch_elastica_s}.}
    \lb{deformate_comp}
    \end{center}
\end{figure}
%%%%%%%%%% END FIG %%%%%%%%%%

Deformed shapes of the elastica (photos taken during tension and compression) at different loadings are reported in Fig.\,\ref{deformate_comp}, while a comparison with theoretical results, denoted with a dashed line, is shown in Fig.\,\ref{sovrapp}. Note the large displacements regime in which the compliant, elastic system operates and the excellent agreement between theory and experiments. 

%%%%%%%%%% FIG %%%%%%%%%%
\begin{figure}[!htcb]
\renewcommand{\figurename}{\footnotesize{Fig.}}
    \begin{center}
    \includegraphics[angle= 0, width = 16 cm]{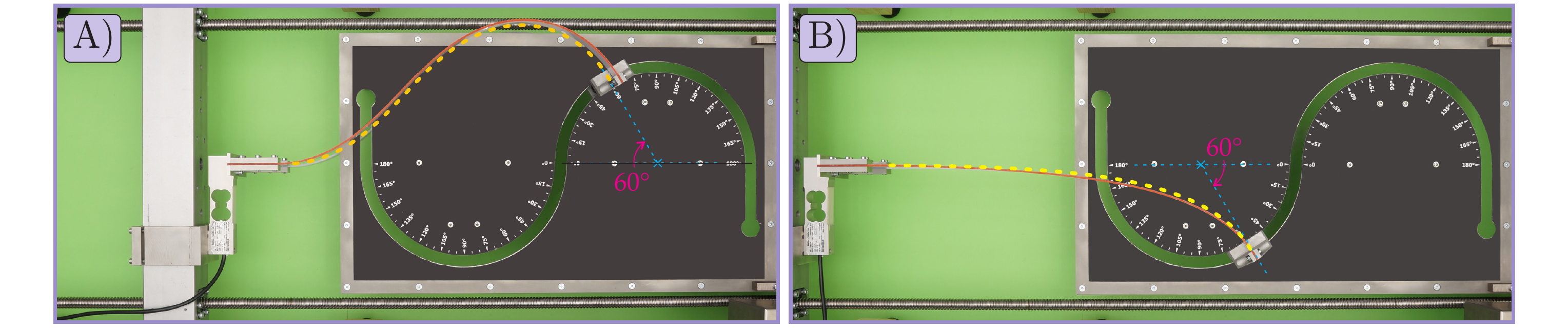}
    \caption{\footnotesize The deformed shape predicted by the elastica  (yellow/dashed line) superimposed on the experimental deformed shapes of the structure (red/solid line). Case (A) refers to compression, while (B) to tension.} 
    \lb{sovrapp}
    \end{center}
\end{figure}
%%%%%%%%%% END FIG %%%%%%%%%%

%
\section*{Conclusions}

A proof-of-concept compliant mechanism has been designed, constructed and tested, in which an elastic rod is constrained at one end with a clamp sliding along a curved and frictionless profile and loaded at the other end. The analyzed constraint introduces nontrivial boundary conditions that strongly affects the elastica, which has been explicitly solved for the mechanical system investigated. Experiments on this system have shown a remarkable agreement with modelling and have opened the possibility of realizing an almost neutral mechanical device, that could be employed as a force limiter.

\section*{Acknowledgments}
D.B., D.M. and D.Z. gratefully acknowledge financial support from the ERC Advanced Grant \lq Instabilities and nonlocal multiscale modelling of materials'
FP7-PEOPLE-IDEAS-ERC-2013-AdG (2014-2019). G.N. acknowledges financial support from SISSA through the excellence program NOFYSAS 2012.

\end{document}